Intrinsic Enhancement of Dielectric Permittivity in (Nb + In) co-doped $TiO_2$ single crystals

Masaru Kawarasaki[1], Kenji Tanabe[1], Ichiro Terasaki[1], and Hiroki Taniguchi[1][*]

[1]Department of Physics, Nagoya University, Nagoya 464-8602, Japan.
[*]e-mail: hiroki_taniguchi@cc.nagoya-u.ac.jp

**The development of dielectric materials with colossal permittivity is important for the miniaturization of electronic devices and fabrication of high-density energy-storage devices. The electron-pinned defect-dipoles has been recently proposed to boost the permittivity of (Nb + In) co-doped $TiO_2$ to $10^5$. However, the follow-up studies suggest an extrinsic contribution to the colossal permittivity from thermally excited carriers. Herein, we demonstrate a marked enhancement in the permittivity of (Nb + In) co-doped $TiO_2$ single crystals at sufficiently low temperatures such that the thermally excited carriers are frozen out and exert no influence on the dielectric response. The results indicate that the permittivity attains quadruple of that for pure $TiO_2$. This finding suggests that the electron-pinned defect-dipoles add an extra dielectric response to that of the $TiO_2$ host matrix. The results offer a novel approach for the development of functional dielectric materials with large permittivity by engineering complex defects into bulk materials.**

In addition to resistors and inductors, capacitors based on the dielectric response of materials are among the most fundamental components of electronic devices. Because capacitance is proportional to the dielectric permittivity ($\varepsilon'$) of the material used, increasing the dielectric permittivity directly improves the density of charge accumulation and thereby supports the development of innovative nanoelectronic and power-electronic devices [1,2]. Several approaches have been proposed for designing large-permittivity materials [3-7]. A standard approach takes advantage of the divergent increase of permittivity in a ferroelectric phase transition [8]. The perovskite-type ferroelectric oxide $BaTiO_3$, which has a cubic-to-tetragonal ferroelectric phase transition at 393 K, has been used as a high-density capacitor in applications requiring permittivities of $\varepsilon' \sim 10^3$ [9]. The effect of nanoscale heterogeneity on the phase transitions of ferroelectrics may push the dielectric permittivity as high as $10^4$ over a wide temperature range in ferroelectrics known as relaxors [10-14]. Quantum fluctuation also enhances the dielectric response by competing with ferroelectric ordering in the low-temperature region, as observed in quantum para/ferroelectrics [15-18]. A colossal permittivity close to $10^5$ has been reported for $CaCu_3Ti_4O_{12}$, which undergoes no ferroelectric phase transition [19]. The origin of this colossal permittivity was traced to the Maxwell–Wagner effect, which stems from the spatial heterogeneity of conductivity in materials due to various reasons, e.g., charge accumulation around the interfaces between the electrodes and materials and/or the grain boundaries in ceramic materials [20-22].

The electron-pinned defect-dipole has recently been proposed as a new route to achieve colossal permittivity by engineering complex defects into bulk materials [23]. This effect has been demonstrated in (Nb + In) co-doped $TiO_2$, whose dielectric permittivity has been shown to reach a value of $6 \times 10^4$ over a frequency range of $10^1$–$10^6$ Hz with little dielectric loss. Studies have suggested that a mutually charge-compensating heterovalent substitution of $Nb^{5+}$ and $In^{3+}$ for $Ti^{4+}$ leads to the formation of complex defects in the $TiO_2$ host matrix; subsequently, the trapped electrons provide the source of the colossal permittivity. However, the follow-up studies have claimed that the colossal permittivity of (Nb + In) co-doped $TiO_2$ has an extrinsic origin in the spatial heterogeneity of conductivity, as in the case of $CaCu_3Ti_4O_{12}$ [24-26].

In this article, we demonstrate an intrinsic enhancement in the dielectric permittivity of (Nb + In) co-doped $TiO_2$ using measurements of the single crystals of $(Nb_{0.5}In_{0.5})_{0.005}Ti_{0.995}O_2$ (NITO-0.5%), whose orientations were confirmed by x-ray diffraction measurements (Fig. 1). The $\varepsilon'$ ~ $10^5$ permittivity of NITO-0.5% was found to decay in the low-temperature region, suggesting that thermally excited electrons accumulate at the sample-electrode interfaces in the high-temperature region and affect the Maxwell–Wagner-type colossal permittivity. We found that at a temperature of approximately 2 K, the permittivity of NITO-0.5% was significantly larger than that of pure $TiO_2$; however, at this temperature, the contribution of the thermally excited electrons is expected to be suppressed completely. The permittivity of NITO-0.5% at this temperature also showed strong anisotropy, with values of $\varepsilon'$ reaching 900 and 250 along the [001] and [110] directions, respectively. The corresponding values for the pure $TiO_2$ were 260 and 120 at this low temperature. Our findings suggest that complex-defect engineering can be used to increase the dielectric permittivity of materials.

Figure 2a shows the real part of the complex dielectric permittivity $\varepsilon'$ of NITO-0.5% measured along the [001] direction and plotted as a function of the measurement frequency and temperature. At 30 K, the highest temperature used in these measurements, the value of $\varepsilon'$ was close to $10^5$ over the entire frequency range of $10^2$–$10^6$ Hz and only little frequency dispersion was observed. Since the measurements were performed on a single crystal, the colossal permittivity originated from the Maxwell–Wagner effect of charge accumulation around the interfaces between the electrodes [26]. The value of $\varepsilon'$ depended strongly on the testing voltage, thereby supporting an extrinsic influence from the interfaces on the dielectric response of NITO-0.5%. The colossal permittivity of NITO-0.5% rapidly disappeared as the temperature decreased, accompanied by a downward shift in the dielectric relaxation frequency, as shown in Fig. 2a. A slowing of the dielectric response with cooling can be clearly seen in the temperature dependence of the characteristic peak shown in Fig. 2b, which represents the frequency dispersion of the imaginary part of the complex permittivity $\varepsilon''$. The activation energy $E_a$ of the thermally excited carriers, which contributes to the colossal permittivity, was estimated to be 7.6 meV using the Arrhenius plot shown in Fig. 2c. This figure shows the logarithmic plot of the peak frequencies of $\varepsilon''$ against the inverse of the temperature.

As can be seen in the low-temperature region of Fig 2a, when the extrinsic colossal permittivity disappears completely, the single crystal of NITO-0.5% still has a permittivity close to $10^3$ along the [001] direction with no dielectric relaxation in the frequency range of $10^2$–$10^6$ Hz. This value is much larger than that of pure TiO$_2$, whose permittivity is ~270 along the [001] direction at low temperatures [27]. As conductive electrons are very weakly excited at 2 K, the marked enhancement of permittivity observed here must have stemmed from the additional electronic states induced by the co-doping of $Nb^{5+}$ and $In^{3+}$. This supports the concept of using electron-pinned defect-dipoles to achieve a large permittivity. Furthermore, the dielectric permittivity of NITO-0.5% in the low-temperature region was found to be strongly anisotropic. The dielectric dispersions measured at 4 K for samples with (001) and (110) wide surfaces are shown by the closed and open circles, respectively, in Fig. 2d. As can be seen, the dielectric permittivity along the [110] direction is 250 over the full frequency range observed herein whereas that along the [001] direction is 900. Moreover, the dielectric permittivity along the [110] axis is much larger than that of pure TiO$_2$ ($\varepsilon'$//[110] ~ 120) [27].

The variation in the dielectric permittivity of NITO-0.5% along the [001] direction at 1 MHz is shown as a function of temperature by the solid circles in Fig. 3. The inset shows a magnified view of the temperature region below 13 K. The temperature dependence of the dielectric permittivity of pure TiO$_2$ along the [001] direction is also plotted in the inset, where open circles denote the results of the present study and squares denote those reported in Ref. 27. The extrinsic colossal permittivity rapidly decays as the temperature decreases before finally disappearing completely at ~13 K. As shown in the inset, under further cooling, the permittivity begins to increase as the temperature decreases. This intrinsic behavior of NITO-0.5% cannot be caused by a Maxwell–Wagner-type dielectric response because the thermally excited electrons freeze at these temperatures. Pure TiO$_2$ is known to show similar behavior across the extended temperature region. We next compare the temperature dependence of the intrinsic dielectric permittivity of NITO-0.5% with that of pure TiO$_2$.

Figure 4 shows the temperature dependence of the inverse permittivity of NITO-0.5% (closed circles) and pure TiO$_2$. The inset shows a magnified view of the temperature region from 0 K to 9 K. Open circles and squares are used to compare the results of the present study with those reported in Ref. 28. The plot with the open squares shows a monotonic decrease in the inverse permittivity of pure TiO$_2$ as the temperature decreases. This suggests an incipient ferroelectricity in pure TiO$_2$, with the system approaching a ferroelectric phase transition as it cools; however, this does not occur in a finite temperature. The virtual phase transition temperature of pure $TiO_2$ can be estimated by extrapolating the temperature dependence of the inverse permittivity using the Curie–Weiss law. The broken line in the figure shows the extrapolation, suggesting a virtual transition temperature of −435 K for pure $TiO_2$. One possible explanation for the permittivity enhancement in NITO-0.5% is that co-doping of $Nb^{5+}$ and $In^{3+}$ shifts the phase transition of $TiO_2$ to a higher temperature and thereby raises the dielectric permittivity. This is often observed in quantum paraelectric and quantum paraelectric-like materials [15-18,29,30]. To investigate this hypothesis, the dielectric permittivity of NITO-0.5% below 13 K was also extrapolated using the Curie–Weiss law. The results are shown in Fig. 4 and its inset. The extrapolation estimated a virtual transition temperature of −19 K, an increase of no less than 400 K. Such a large shift seems implausible for the amount of dopant present in NITO-0.5%.

Thus, the most plausible origin of the large permittivity of NITO-0.5% is the response of the electron-pinned defect-dipoles, as originally proposed in Ref. 23. Assuming that the difference in the permittivity between the NITO-0.5% and pure $TiO_2$ at 4 K, i.e., $\Delta\varepsilon' = 640$, stems from the response of the electron-pinned defect-dipoles, an induced polarization of $5.66 \times 10^{-5}$ $Cm^{-2}$ will be required at the present testing voltage of 100 V/cm. The defect density of NITO-0.5%, assuming one complex defect every 200 unit cells, is estimated to be $\sim 1.6 \times 10^{20}$ $cm^{-3}$. An averaged dipole moment of $3.54 \times 10^{-31}$ Cm for a single complex defect is then estimated from the value of the induced polarization. The induced dipole moment estimated here corresponds to a reduced displacement of the elementary charge of ~0.02 Å under a voltage of 100 V/cm. This estimated reduced displacement is equivalent to several percents of the size of a complex defect, which is considered to spread over few Å [23]. To increase the complex-defect-induced permittivity further, the defect density must be increased and/or the defect state should be made dispersive. However, such modifications would require a tradeoff with the insulating properties of the material. A next step would be to optimize the defect-induced dielectric response via the design of the electronic states in the complex defects.

Although further quantitative investigation is required, the results of the present study demonstrate that co-doping of $Nb^{5+}$ and $In^{3+}$ introduces an additional agent that responds to the external electric field in the host $TiO_2$ matrix and thereby enhances its intrinsic dielectric permittivity. We believe that this work elucidates the potential of defect engineering in increasing the dielectric permittivity of a host matrix and contributing to the development of innovative nanoelectronic and power-electronic devices.

**Methods**

Stoichiometric mixtures of $TiO_2$(4N), $In_2O_3$(3N), and $Nb_2O_5$(4N) were subjected to hydrostatic pressing to produce rods in crystal growth. These rods were then sintered at 1673 K for 10 h in a box furnace. Single crystals were grown using an optical floating zone furnace (FZ-T-4000-H-I-V;

Crystal Systems) equipped with a single ellipsoidal gold-plated mirror. The obtained single crystals of NITO-0.5% were sliced using a diamond cutter to prepare two samples with (001) and (110) wide surfaces. The (001) and (110) orientations of the obtained samples were confirmed using X-ray diffraction (RIGAKU RINT-2000), as shown in Fig. 1. Dielectric permittivity was measured using a Quantum Design Physical Property Measurement System equipped with a Keysight 4284A precision LCR-meter in the temperature region from 2 K to 30 K.

**References**

1. Homes, C. C., Vogt, T., Shapiro, S. M., Wakimoto, S. & Ramirez, A. P. Optical response of high-dielectric-constant perovskite-related oxide. *Science* **293**, 673-676 (2001).

2. Krohns, S., Lunkenheimer, P., Meissner, S., Reller, A., Gleich, B., Rathgeber, A., Gaugler, T., Buhl, H. U., Sinclair, D. C. & Loidl, A. The route to resource-efficient novel materials. *Nature Mater.* **10**, 899-901 (2011).

3. Heywang, W. Semiconducting Barium Titanate. *J. Mater. Sci.* **6**, 1214-1226 (1971).

4. Krohns, S., Lunkenheimer, P., Kant, Ch., Pronin, A. V., Brom, H. B., Nugroho, A. A., Diantoro, M. & Loidl, A. Colossal dielectric constant up to gigahertz at room temperature. *Appl. Phys. Lett.* **94**, 122903 (2009).

5. Shu-Nan, M. & Maki, M. Dielectric anisotropy in the charge-density-wave state of $K_{0.3}MoO_3$. *J. Phys. Soc. Jpn.* **80**, 084706 (2011).

6. Wu, J. B., Nan, C. W., Lin, Y. H. & Deng, Y. Giant dielectric permittivity observed in Li and Ti doped NiO. *Phys. Rev. Lett.* **89**, 217601 (2002).

7. Hess, H. F., Deconde, K., Rosenbaum, T. F. & Thomas, G. A. Giant dielectric constants at the approach to the insulator-metal transition. *Phys. Rev. B* **25**, 5578-5580 (1982).

8. Lines, M. E. & Glass, A. M. *Principles and applications of ferroelectrics and related materials* (Clarendon Press, Oxford, 1977).

9. Merz, W. J. The electric and optical behavior of $BaTiO_3$ single-domain crystals. *Phys. Rev.* **76**, 1221 (1949).

10. Cross, L. E. Relaxor ferroelectrics. *Ferroelectrics* **76**, 241-267 (1987).

11. Burns, G. & Dacol, F. H. Crystalline ferroelectrics with glassy polarization behavior. *Phys. Rev. B* **28**, 2527-2530 (1983).

12. Fu, D., Taniguchi, H., Itoh, M., Koshihara, S., Yamamoto, N. & Mori, S. Relaxor $Pb(Mg_{1/3}Nb_{2/3})O_3$: a ferroelectric with multiple inhomogeneities. *Phys. Rev. Lett.* **103**, 207601 (2009).

13. Xu, G., Wen, J., Stock, C. & Gehring, P. M. Phase instability induced by polar nanoregions in a relaxor ferroelectric system. *Nature Mater.* **7**, 562-566 (2008).

14. Manley, M. E., Lynn, J. W., Abernathy, D. L., Specht, E. D., Delaire, O., Bishop, A. R., Sahul, R. & Budai, J. D. Phonon localization drives polar nanoregions in a relaxor ferroelectric. *Nature Commun.* **5**, 3683 (2014).

15. Müller, K. A. & Burkard, H. $SrTiO_3$: An intrinsic quantum paraelectric below 4 K. *Phys. Rev. B* **19**, 3593 (1979).

16. Müller, K. A. & Bednorz, J. G. $Sr_{1-x}Ca_xTiO_3$: An XY quantum ferroelectric with transition to randomness. *Phys. Rev. Lett.* **52**, 2289 (1984).

17. Itoh, M., Wang, R., Inaguma, Y., Yamaguchi, T., Shan, Y-J. & Nakamura, T. Ferroelectricity induced by oxygen isotope exchange in strontium titanate perovskite. *Phys. Rev. Lett.* **82**, 3540 (1999).

18. Taniguchi, H., Itoh, M. & Yagi, T. Ideal soft mode-type quantum phase transition and phase coexistence at quantum critical point in $^{18}$O-Exchanged $SrTiO_3$ *Phys. Rev. Lett.* **99**, 017602 (2007).

19. Subramanian, M. A., Li, D.; Duan, N., Reisner, B. A. & Sleight, A. W. High dielectric constant in $ACu_3Ti_4O_{12}$ and $ACu_3Ti_3FeO_{12}$ phases. *J. Solid State Chem.* **151**, 323–325 (2000).

20. Wang, C. C. & Zhang, L. W. Surface-layer effect in $CaCu_3Ti_4O_{12}$. *Appl. Phys. Lett.* **88**, 042906 (2006).

21. Fu, D., Taniguchi, H., Taniyama, T., Itoh, M. & Koshihara, S. Origin of giant dielectric response in nonferroelectric $CaCu_3Ti_4O_{12}$: Inhomogeneous conduction nature probed by atomic force microscopy. *Chem. Mater.* **20**, 1694-1698 (2008).

22. Chiou, B-S., Lin, S-T., Duh, J-G. & Chang, P-H. Equivalent circuit model in grain-boundary barrier layer capacitors. *J. Am. Ceram. Soc.* **72**, 1967-1975 (1989).

23. Hu, W., Liu, Y., Withers, R. L., Frankcombe, T. J., Norén, L., Snashall, A., Kitchin, M., Smith,P., Gong, B., Chen, H., Schiemer, J., Brink, F. & Wong-Leung, J. Electron-pinned defect-dipoles for high-performance colossal permittivity materials. *Nature Mater.* **12**, 821-826 (2013).

24. Li, J., Li, F., Li, C., Yang, G., Xu, Z. & Zhang, S. Evidences of grain boundary capacitance effect on the colossal dielectric permittivity in (Nb+In) co-doped $TiO_2$ ceramics. *Sci. Rep.* **5**, 8295 (2015).

25. Mandal, S., Pal, S., Kundu, A. K., Menon, K. S. R., Hazarika, A., Rioult, M. & Belkhou, R. Direct view at colossal permittivity in donor-acceptor (Nb, In) co-doped rutile $TiO_2$. *Appl. Phys. Lett.* **109**, 092906 (2016).

26. Song, Y., Wang, X., Sui, Y., Liu, Z., Zhang, Y., Zhan, H., Song, B., Liu, Z., Lv, Z., Tao, L. & Tang J. Origin of colossal dielectric permittivity of rutile $Ti_{0.9}In_{0.05}Nb_{0.05}O_2$: single crystal and polycrystalline *Sci. Rep.* **6**, 21478 (2016).

27. Parker, R. A. Static dielectric constant of rutile ($TiO_2$), 1.6-1060°K. *Phys. Rev.* **124**, 1719 (1961).

28. Samara, G. A. & Peercy, P. S. Pressure and temperature dependence of the static dielectric

constants and Raman spectra of TiO$_2$ (Rutile). *Phys. Rev. B* **7**, 1131 (1973).

29. Rytz, D., Höchli, U. T. & Bilz, H. Dielectric susceptibility in quantum ferroelectrics. *Phys. Rev. B* **22**, 359 (1980).

30. Lemanov, V. V., Sotnikov, A. V., Smirnova, E. P. & Weihnacht, M. From incipient ferroelectricity in CaTiO$_3$ to real ferroelectricity in Ca$_{1-x}$Pb$_x$TiO$_3$ solid solutions. *Appl. Phys. Lett.* **81**, 886 (2002).

**Acknowledgements**

This work is partially supported by a Grant-in-Aid for Young Scientists (A) (No.16H06115), MEXT Element Strategy Initiative Project.

**Figure Captions**

Fig. 1: X-ray diffraction patterns of NITO-0.5% single crystals with (a) (001) and (b) (110) wide surfaces, wherein the incident X-ray was irradiated normal to the wide surface. The calculated diffraction pattern of TiO$_2$ is presented in Panel (c).

Fig. 2: Frequency dispersions for (a) the real and (b) imaginary parts of the dielectric permittivity of a NITO-0.5% single crystal with a (001) wide surface measured as a function of temperature ranging from 2 K to 30 K. Panel (c) shows an Arrhenius plot for the relaxation frequency against the inverse temperature. The activation energy of 7.6 meV for the thermally excited carriers was estimated from linear fitting and is shown by the solid line. Panel (d) shows the frequency dispersions for the dielectric permittivity in NITO-0.5% at 4 K measured along the [001] and the [110] directions, which are represented by closed and open circles, respectively.

Fig. 3: Real part of the dielectric permittivity of a NITO-0.5% single crystal with a (001) wide surface measured at 1 MHz and shown as a function of temperature by closed circles. The inset shows a magnified view of the temperature region below 13 K, where the open circles and squares indicate the real part of the dielectric permittivity of pure TiO2 observed in the present study and that reported in [27].

Fig. 4: Inverse permittivity of a NITO-0.5% single crystal with a (001) wide surface measured at 1 MHz and plotted as a function of temperature by closed circles. Open circles and squares denote the results for pure TiO2 obtained in the present study and those reported in [28]. The solid and broken lines indicate the extrapolated temperature dependences of NITO-0.5% and pure TiO2. These were obtained by fitting the linear parts of the plots using the Curie–Weiss law to estimate the virtual phase transition temperatures. The inset shows a magnified view of the temperature region below 9 K.

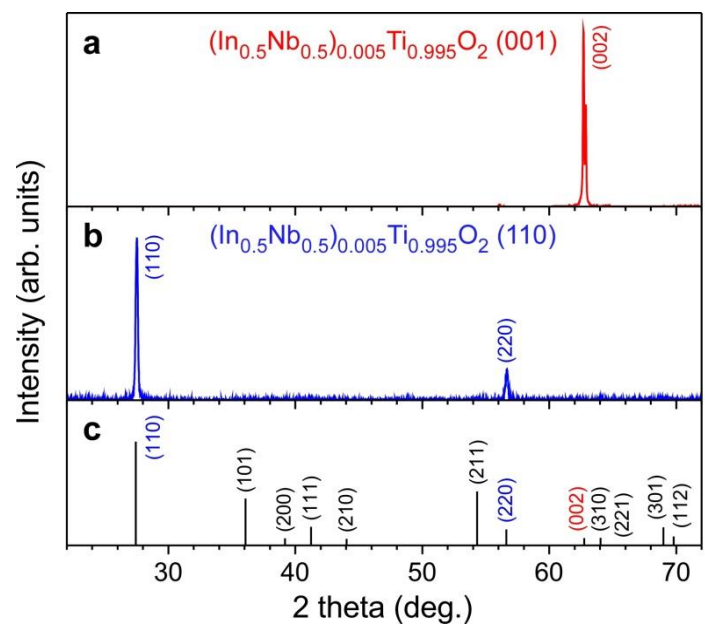

Figure 1, Kawarasaki *et al*.

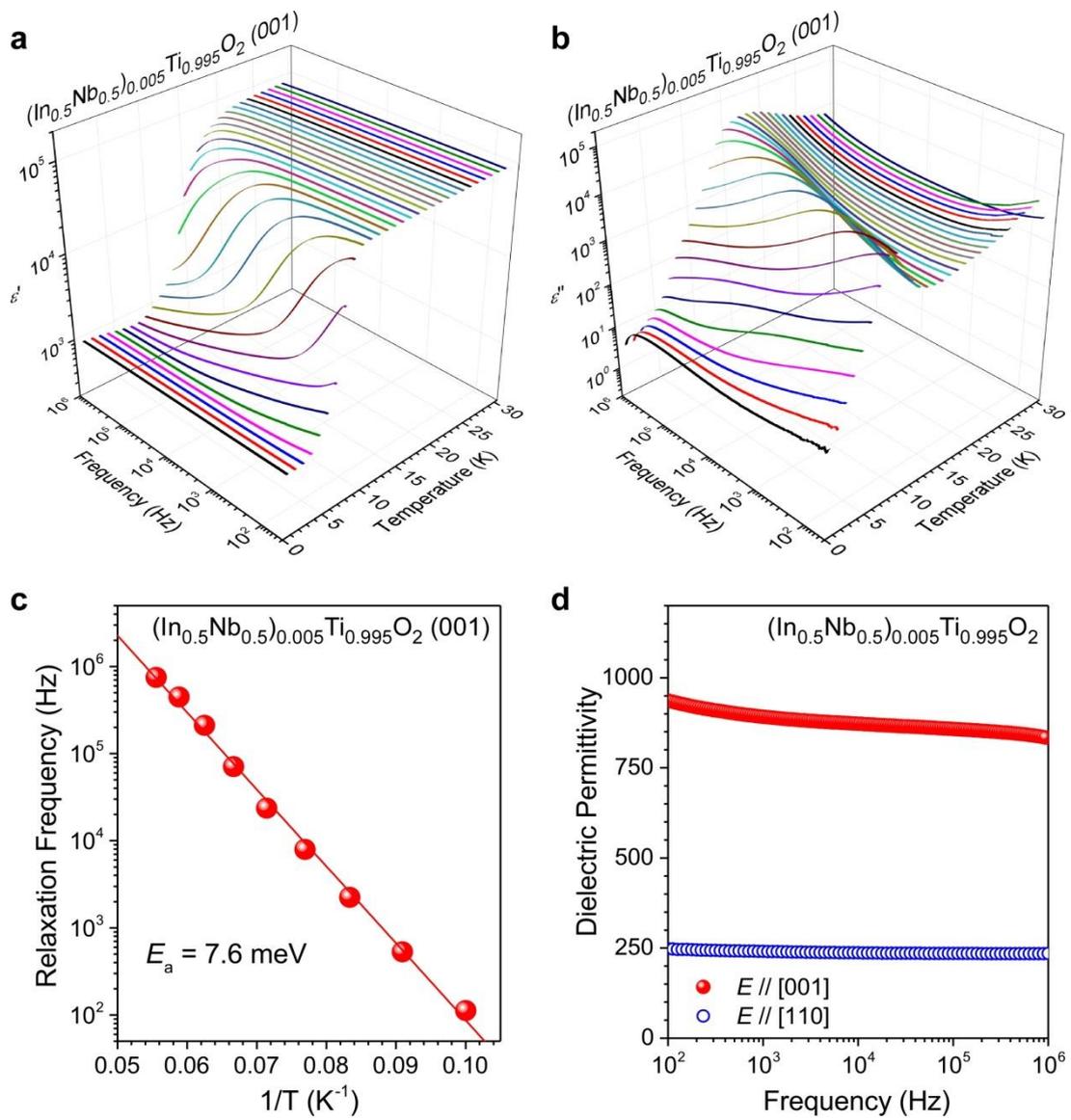

Figure 2, Kawarasaki *et al*.

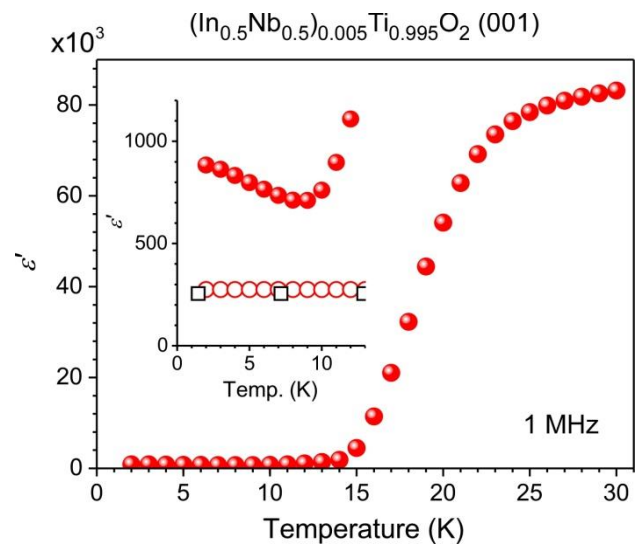

Figure 3, Kawarasaki *et al*.

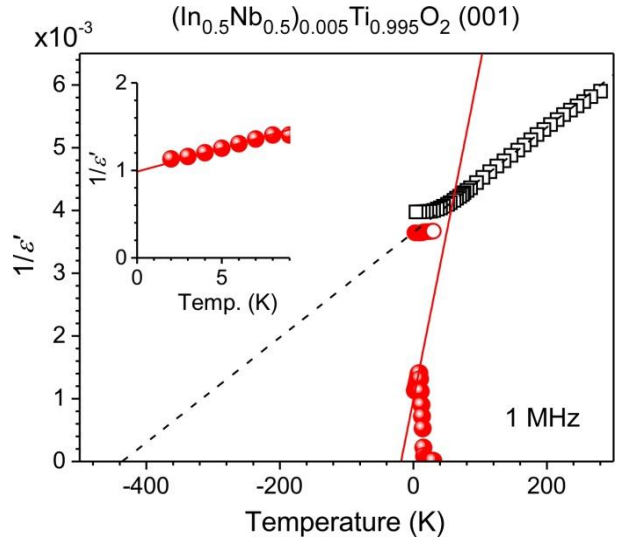

Figure 4, Kawarasaki *et al*.

Dear Editor:

Please find enclosed our manuscript entitled "Intrinsic Enhancement of Dielectric Permittivity in (Nb + In) co-doped $TiO_2$ single crystals," which we request you to consider for publication as an *Article* in *Nature Materials*.

Materials with large dielectric permittivity have long been explored because they are extremely important for the miniaturization of electronic devices and development of high-density energy-storage devices. These materials are conventionally developed by engineering ferroelectric phase transitions and spatial heterogeneity of conductive properties in materials. Recently, the complex-defect-engineering involving a mutually charge-compensating heterovalent substitution was proposed as a new approach for achieving large permittivity. This has attracted keen attention because the dielectric permittivity of the materials designed by this strategy reaches $\varepsilon' \sim 10^6$, with only a small dielectric loss over a wide temperature range. However, the follow up studies indicate extrinsic contributions of thermally excited carriers to the colossal permittivity and have provoked a controversy over the validity of the proposed strategy.

In this paper, we show that the intrinsic enhancement in the permittivity of (Nb + In) co-doped $TiO_2$ is caused by the dielectric response of complex defects. This result is supported by the dielectric measurements of single crystals in a low-temperature region, wherein the undesirable effects of thermally excited electrons are completely eliminated. The present study elucidates the potential of defect engineering as a novel strategy for boosting the dielectric permittivity of a host matrix and developing innovative nanoelectronic and power-electronic devices.

We believe that the present study is relevant to the scope of *Nature Materials* and meets the criteria for publication.

Sincerely,

Hiroki Taniguchi
Department of Physics,
Nagoya University,
Nagoya 464-8602,
Japan.
Email Address: hiroki_taniguchi@cc.nagoya-u.ac.jp